\documentclass[aps]{revtex4}
\usepackage{amsmath}
\usepackage{mathrsfs}
\usepackage{bm}

\begin{document}

\title{Nonlinear interaction between three inertial Alfv\'{e}n waves. }
\author{G. Brodin, L. Stenflo and P. K. Shukla \\
Department of Physics, Ume\aa\ University, SE-901 87 Ume\aa , Sweden}

\begin{abstract}
The resonant coupling between Alfv\'{e}n waves is reconsidered. New results
are found for cold plasmas there temperature effects are negligible. 
\end{abstract}

\maketitle

\draft


The nonlinear interaction between Alfv\'{e}n waves (Sagdeev and Galeev,
1969; Hasegawa and Uberoi, 1982; Petviashvili and Pokhotelov, 1992) is a key
subject within modern plasma science (Shukla, 2004) that is still a matter
of subtle analysis (e.g. Fedun et al. 2004; Voitenko and Goossens, 2005;
Shukla and Stenflo, 2005). In order to put the theory on a firm theoretical
basis, we shall in the present paper consider the resonant interaction
between three inertial Alfv\'{e}n waves in a uniform \textit{cold}
magnetized plasma, starting our analysis from the \textit{exact} expressions
for the coupling coefficients in such a plasma. Considering the appropriate
frequency limits we shall then derive results which have not been presented
previously.

\ Let us thus investigate the resonant interaction between three waves with
frequencies $\omega _{j}$ ($j=1,2,3$) and wave vectors $\mathbf{k}_{j}$, and
assume that the matching conditions 
\begin{equation}
\omega _{3}=\omega _{1}+\omega _{2}  \label{frequency}
\end{equation}
and 
\begin{equation}
\mathbf{k}_{3}=\mathbf{k}_{1}+\mathbf{k}_{2}  \label{wave-vector}
\end{equation}
are satisfied. The development of, for example, the z-components ($E_{jz})$
of the wave electric field amplitudes is then governed by the three coupled
bilinear equations (e.g. Stenflo, 1994) 
\begin{mathletters}
\begin{equation}
\frac{dE_{1z}^{\ast }}{dt}=\alpha _{1}E_{2z}E_{3z}^{\ast }  \tag{3a}
\end{equation}
\begin{equation}
\frac{dE_{2z}^{\ast }}{dt}=\alpha _{2}E_{1z}E_{3z}^{\ast }%
\addtocounter{equation}{2}  \tag{3b}
\end{equation}
and 
\begin{equation}
\frac{dE_{3z}}{dt}=\alpha _{3}E_{2z}E_{1z}  \tag{3c}
\end{equation}
where the z-axis is along the external magnetic field ($B_{0}\widehat{%
\mathbf{z}}$), the star denotes complex conjugate, $\alpha _{j}$ are the
coupling coefficients, $d/dt=\partial /\partial t+\mathbf{v}_{gj}\cdot
\nabla +\nu _{j}$ where $\mathbf{v}_{gj}$ is the group velocity of wave $j$,
and $\nu _{j}$ accounts for the linear damping rate. As shown by Stenflo and
Brodin (2005) the coefficients $\alpha _{j}$ for a cold plasma are 
\end{mathletters}
\begin{equation}
\alpha _{1,2}=\frac{M_{1,2}}{\partial D(\omega _{1,2},\mathbf{k}%
_{1,2})/\partial \omega _{1,2}}C  \tag{4a,b}
\end{equation}
and 
\begin{equation}
\alpha _{3}=-\frac{M_{3}}{\partial D(\omega _{3},\mathbf{k}_{3})/\partial
\omega _{3}}C  \tag{4c}
\end{equation}
where 
\begin{eqnarray}
C &=&\sum_{\sigma }\frac{q\omega _{p}^{2}}{m\omega _{1}\omega _{2}\omega
_{3}k_{1z}k_{2z}k_{3z}}\left[ \frac{\mathbf{k}_{1}\mathbf{\cdot K}_{1}}{%
\omega _{1}}\mathbf{K}_{2}\mathbf{\cdot K}_{3}^{\ast }+\frac{\mathbf{k}_{2}%
\mathbf{\cdot K}_{2}}{\omega _{2}}\mathbf{K}_{1}\mathbf{\cdot K}_{3}^{\ast }+%
\frac{\mathbf{k}_{3}\mathbf{\cdot K}_{3}^{\ast }}{\omega _{3}}\mathbf{K}_{1}%
\mathbf{\cdot K}_{2}-\right.  \notag \\
&&\left. \frac{i\omega _{c}}{\omega _{3}}\left( \frac{k_{2z}}{\omega _{2}}-%
\frac{k_{1z}}{\omega _{1}}\right) \mathbf{K}_{3}^{\ast }\mathbf{\cdot }%
\left( \mathbf{K}_{1}\times \mathbf{K}_{2}\right) \right]  \label{C-coeff}
\end{eqnarray}
\begin{equation}
\mathbf{K}=-\left[ \mathbf{k}_{\bot }+i\frac{\omega _{c}}{\omega }\mathbf{k}%
\times \widehat{\mathbf{z}}+\left( \frac{\sum i\frac{\omega _{c}}{\omega }%
\frac{\omega _{p}^{2}}{\omega ^{2}-\omega _{c}^{2}}}{1-\frac{k^{2}c^{2}}{%
\omega ^{2}}-\sum \frac{\omega _{p}^{2}}{\omega ^{2}-\omega _{c}^{2}}}%
\right) \left( \mathbf{k}\times \widehat{\mathbf{z}}-i\frac{\omega _{c}}{%
\omega }\mathbf{k}_{\bot }\right) \right] \frac{\left( 1-\frac{k_{\bot
}^{2}c^{2}}{\omega ^{2}}-\sum \frac{\omega _{p}^{2}}{\omega ^{2}}\right)
\omega ^{4}}{\left( \omega ^{2}-\omega _{c}^{2}\right) k_{\bot }^{2}c^{2}}%
+k_{z}\widehat{\mathbf{z}}  \label{k-coeff-vector}
\end{equation}

\begin{eqnarray}
D(\omega ,\mathbf{k}) &=&\left( 1-\frac{k^{2}c^{2}}{\omega ^{2}}-\sum \frac{%
\omega _{p}^{2}}{\omega ^{2}-\omega _{c}^{2}}\right) \left[ \left( (1-\frac{%
k_{z}^{2}c^{2}}{\omega ^{2}}-\sum \frac{\omega _{p}^{2}}{\omega ^{2}-\omega
_{c}^{2}}\right) \left( 1-\frac{k_{\bot }^{2}c^{2}}{\omega ^{2}}-\sum \frac{%
\omega _{p}^{2}}{\omega ^{2}}\right) -\right.  \notag \\
&&\left. \frac{k_{\bot }^{2}k_{z}^{2}c^{4}}{\omega ^{4}}\right] -\left( \sum 
\frac{\omega _{p}^{2}\omega _{c}}{\omega (\omega ^{2}-\omega _{c}^{2})}%
\right) ^{2}\left( 1-\frac{k_{\bot }^{2}c^{2}}{\omega ^{2}}-\sum \frac{%
\omega _{p}^{2}}{\omega ^{2}}\right)  \label{DR}
\end{eqnarray}
and 
\begin{equation}
M_{j}=\left( 1-\frac{k_{j}^{2}c^{2}}{\omega _{j}^{2}}-\sum \frac{\omega
_{p}^{2}}{\omega _{j}^{2}-\omega _{c}^{2}}\right) \left( 1-\frac{%
k_{jz}^{2}c^{2}}{\omega _{j}^{2}}-\sum \frac{\omega _{p}^{2}}{\omega
_{j}^{2}-\omega _{c}^{2}}\right) -\left( \sum \frac{\omega _{p}^{2}\omega
_{c}}{\omega _{j}(\omega _{j}^{2}-\omega _{c}^{2})}\right) ^{2}
\label{M-formula}
\end{equation}
where $k=(k_{z}^{2}+k_{\bot }^{2})^{1/2}$, $\mathbf{k}_{\bot }$ is the
perpendicular (to $\widehat{\mathbf{z}}$) part of the wave-vector, $\omega
_{p}$ is the plasma frequency ($\omega _{pe}$ for the electrons and $\omega
_{pi}$ for the ions), $\omega _{c}=qB_{0}/m$ is the cyclotron frequency, $q$
and $m$ are the particle charge and mass, and $c$ is the speed of light in
vacuum. For notational convenience, the subscript $\sigma $ denoting the
various particle species has been left out in the above formulas.

Equations (3)-(8) can be used to study the high-frequency coherent
generation (Christiansen et al., 1981) or the energy transfer from a large
amplitude electromagnetic wave into an electrostatic electron wave and a
lower hybrid wave (e.g. Larsson et al., 1976; Stenflo, 1994; Kuo 2003;
Stenflo, 2004). In the present paper we shall, however, only adopt the above
results to investigate wave couplings in the MHD regime. In a previous work
(Brodin and Stenflo, 1988) we noted that in the ideal MHD regime, there is
no coupling between three shear Alfv\'{e}n waves. We will therefore show
below that the more accurate two fluid model gives predictions significantly
different from ideal MHD. \ To demonstrate this fact, we will consider
low-frequency waves where the parameters have the standard MHD ordering $%
\omega \sim k_{z}V_{A}\ll \omega _{ci}$ and $\omega _{ci}\ll \omega _{pi}$,
where $V_{A}$ is the Alfv\'{e}n velocity. However, in contrast to ideal MHD,
we shall here also allow for large perpendicular wave numbers, reaching up
to $k_{\bot }\sim \omega _{pe}/c$. Applying this scaling to Eq. (\ref{DR}),
we find that the usual shear Alfv\'{e}n waves are modified to inertial
Alfv\'{e}n waves, with frequencies 
\begin{equation}
\omega \simeq \frac{k_{z}V_{A}}{1+k_{\bot }^{2}\lambda _{e}^{2}}
\label{IAW-disp}
\end{equation}
where $\lambda _{e}=c/\omega _{pe}$. Next we assume that all the three
interacting modes are described by (\ref{IAW-disp}). In order to keep
contact with the ideal MHD regime where $\omega \simeq k_{z}V_{A}$, we will
allow for $k_{\bot }^{2}\lambda _{e}^{2}\ll 1$ as well as for $k_{\bot
}^{2}\lambda _{e}^{2}\sim 1$. Applying the above assumptions to (\ref
{k-coeff-vector}), we thus make the approximations 
\begin{equation}
\mathbf{K}_{e}\mathbf{\simeq }-\frac{i\omega }{\omega _{ce}}\frac{(1+k_{\bot
}^{2}\lambda _{e}^{2})}{k_{\bot }^{2}\lambda _{e}^{2}}\mathbf{k}\times 
\widehat{\mathbf{z}}+k_{z}\widehat{\mathbf{z}}  \label{K-electron}
\end{equation}
and 
\begin{equation}
\mathbf{K}_{i}\mathbf{\simeq }-\frac{i\omega }{\omega _{ci}}\frac{(1+k_{\bot
}^{2}\lambda _{e}^{2})}{k_{\bot }^{2}\lambda _{e}^{2}}\mathbf{k}\times 
\widehat{\mathbf{z}}+k_{z}\widehat{\mathbf{z}}  \label{K-ion}
\end{equation}
Substituting (\ref{K-electron}) and (\ref{K-ion}) into (\ref{C-coeff}), we
find that the ion contribution is negligible compared to the electron
contribution, and that the coupling coefficient reduces to 
\begin{equation}
C_{IAW}=\frac{q_{e}\omega _{pe}^{2}}{m_{e}\omega _{1}\omega _{2}\omega _{3}}%
\left( \frac{k_{1z}}{\omega _{1}}+\frac{k_{2z}}{\omega _{2}}+\frac{k_{3z}}{%
\omega _{3}}\right)  \label{C-approx}
\end{equation}
and that the dispersion function (\ref{DR}) can be approximated by 
\begin{equation}
D(\omega ,\mathbf{k})=-\frac{c^{4}\left( \omega ^{2}-k^{2}V_{A}^{2}\right)
\left( k_{\bot }^{2}c^{2}+\omega _{pe}^{2}\right) }{\omega ^{6}V_{A}^{4}}%
\left[ \omega ^{2}-\frac{k_{z}^{2}V_{A}^{2}}{1+k_{\bot }^{2}c^{2}/\omega
_{pe}^{2}}\right] .  \label{D-IAW}
\end{equation}
Similarly (\ref{M-formula}) reduces to 
\begin{equation}
M=\frac{c^{4}}{\omega ^{4}V_{A}^{4}}\left( \omega ^{2}-k^{2}V_{A}^{2}\right)
\left( \omega ^{2}-k_{z}^{2}V_{A}^{2}\right) .  \label{M-IAW}
\end{equation}
Thus 
\begin{equation}
\alpha _{1,2}=\frac{\omega _{1,2}^{3}k_{1,2\bot }^{2}c^{2}}{2\left(
k_{1,2\bot }^{2}c^{2}+\omega _{pe}^{2}\right) \omega _{pe}^{2}}C_{IAW}
\label{alfa-IAW}
\end{equation}
and 
\begin{equation}
\alpha _{3}=-\frac{\omega _{3}^{3}k_{3\bot }^{2}c^{2}}{2\left( k_{3\bot
}^{2}c^{2}+\omega _{pe}^{2}\right) \omega _{pe}^{2}}C_{IAW}  \label{alfa3}
\end{equation}
For easy comparison with the MHD-results, it is convenient to eliminate $%
E_{z},$ and to work with the magnetic field amplitudes. The magnitude of the
magnetic field perturbation is thus 
\begin{equation}
B_{j}=\frac{E_{jz}}{\omega k_{j\bot }\lambda _{e}^{2}}  \label{BE-relation}
\end{equation}
where the magnetic field is directed in the $-\mathbf{k}\times \widehat{%
\mathbf{z}}$-direction, to a good approximation. The coupled equations
(3,a,b,c) can therefore be rewritten as 
\begin{equation}
\frac{dB_{1,2}}{dt}=\frac{\omega _{1,2}k_{1\bot }k_{2\bot }k_{3\bot
}V_{A}^{2}}{2\left( k_{1,2\bot }^{2}+\lambda _{e}^{-2}\right) \omega _{ci}}%
\left( \frac{k_{1z}}{\omega _{1}}+\frac{k_{2z}}{\omega _{2}}+\frac{k_{3z}}{%
\omega _{3}}\right) \frac{B_{1,2}^{\ast }B_{3}}{B_{0}}  \label{B12-coupling}
\end{equation}
and 
\begin{equation}
\frac{dB_{3}}{dt}=-\frac{\omega _{3}k_{1\bot }k_{2\bot }k_{3\bot }V_{A}^{2}}{%
2\left( k_{3\bot }^{2}+\lambda _{e}^{-2}\right) \omega _{ci}}\left( \frac{%
k_{1z}}{\omega _{1}}+\frac{k_{2z}}{\omega _{2}}+\frac{k_{3z}}{\omega _{3}}%
\right) \frac{B_{1}B_{2}}{B_{0}}  \label{B3-coupling}
\end{equation}
Chosing wave 3 as a pump wave, the maximum growth rate $\gamma _{\max }$
deduced from (\ref{B12-coupling}) is thus of the order 
\begin{equation}
\gamma _{\max }\sim \frac{\omega _{3}k_{3\bot }V_{A}}{\omega _{ci}}\frac{%
\left| B_{3}\right| }{B_{0}}  \label{g-max1}
\end{equation}
where the fastest growth occurs for decay products that are inertial
Alfv\'{e}n waves with wavenumbers $k_{1,2\bot }$ of the order of $\lambda
_{e}^{-1}$. We note that for decay into standard ideal Alfv\'{e}n waves with 
$k_{1,2\bot }\sim k_{1,2z}=\omega _{1,2}/V_{A}$, the growth rate thus is
reduced by a factor $\sim \omega _{3}^{2}m_{e}/\omega _{ci}^{2}m_{i}$. As a
special limit of (\ref{g-max1}) we consider a pump wave which is a standard
ideal Alfv\'{e}n wave with $k_{3\bot }\sim k_{3z}=\omega _{3}/V_{A}$. This
gives 
\begin{equation}
\gamma _{\max }\sim \frac{\omega _{3}^{2}}{\omega _{ci}}\frac{\left|
B_{3}\right| }{B_{0}}  \label{g-max2}
\end{equation}
As another example, we let the pump wave be an inertial Alfven pump wave
with perpendicular wavenumber $k_{3\bot }\sim \lambda _{e}^{-1}$, in which
case (\ref{g-max1}) reduces to 
\begin{equation}
\gamma _{\max }\sim \omega _{3}\frac{\left| B_{3}\right| }{B_{0}}\left( 
\frac{m_{i}}{m_{e}}\right) ^{1/2}  \label{g-max3}
\end{equation}

To summarize, we have considered the interaction of Alfv\'{e}n waves using
results from the exact two-fluid equations for a cold magnetized plasma. We
note that for an Alfv\'{e}n pump wave in the inertial regime ($k_{3\bot
}\sim \lambda _{e}^{-1}$), the maximum growth rate for decay into inertial
Alfv\'{e}n waves is larger than the usual MHD growth rates (including
interaction between all sorts of ideal MHD waves) by a factor of the order
of $(m_{i}/m_{e})^{1/2}$. Furthermore, for an ordinary Alfv\'{e}n pump wave (%
$k_{3\bot }\ll \lambda _{e}^{-1}$), we point out that the cold ideal\ MHD
theory does not allow for resonant decay processes at all (Brodin and
Stenflo, 1988), since the Manley-Rowe relations prevent decay into modes of
higher frequencies, and the coupling coefficients for interaction with two
other Alfven waves are zero in the framework of ideal MHD. Thus the growth
rate found in (\ref{g-max2}) is the fastest decay possible in a cold plasma
for an Alfv\'{e}n pump wave in the ideal MHD regime ($k_{3\bot }\ll \lambda
_{e}^{-1}$). \ Furthermore, since the decay products have short scale
lengths, we note that the ideal\ MHD equations are unable to describe the 
\textit{nonlinear} evolution, even if the initial conditions lie well inside
the usual validity conditions of those equations. We point out that
resisitivity eventually leads to dissipation of the shorter scale waves.
Thus we conclude that the parametric processes considered in this paper can
be important for understanding the heating of low-beta plasmas. Finally we
stress that high-beta plasmas require a separate analysis (Brodin and
Stenflo, 1990) and that extensions to three-wave interactions in a turbulent
plasma (Vladimirov and Yu, 2004) are comparatively straightforward.

\end{document}